\documentclass[twocolumn,notitlepage,superscriptaddress]{revtex4-1} 
\usepackage[colorlinks=true,urlcolor=blue,linkcolor=blue,citecolor=blue]{hyperref}
\usepackage{latexsym}
\usepackage{graphicx}
\usepackage{subfigure} 
\usepackage{placeins}
\usepackage{amssymb}
\usepackage{amsmath}
\usepackage{color}
\usepackage{amsfonts}
\usepackage{mathrsfs}
\usepackage{simpler-wick}

\providecommand \doibase [0]{http://dx.doi.org/} 

\usepackage{accents}
\makeatletter
\def\wideubar{\underaccent{{\cc@style\underline{\mskip10mu}}}}
\def\Wideubar{\underaccent{{\cc@style\underline{\mskip8mu}}}}
\makeatother
\makeatletter
\def\widebar{\accentset{{\cc@style\underline{\mskip10mu}}}}
\def\Widebar{\accentset{{\cc@style\underline{\mskip8mu}}}}
\makeatother

\begin{document}

\title{Microscopic Amp\`ere  current-current interaction }

\author{Yuehua Su}
\email{suyh@ytu.edu.cn}
\affiliation{ Department of Physics, Yantai University, Yantai 264005, People's Republic of China }

\author{Desheng Wang}
\affiliation{ Department of Physics, Yantai University, Yantai 264005, People's Republic of China }

\author{Chao Zhang}
\affiliation{ Department of Physics, Yantai University, Yantai 264005, People's Republic of China }

\begin{abstract}

With the rapid development of modern measurement techniques, the energy resolution of $1 \, meV$ can now be easily obtained. Generally, the driving mechanisms of the physical, chemical or biological processes of the matters or the living organisms on Earth at about $1 \, meV$ energy scale are assumed to stem from the fundamental microscopic Coulomb interaction, its various reduced ones and the relativistic corrections. In this article, by using a path integral approach on a non-relativistic quantum electrodynamics theory, we show that there is another fundamental microscopic electromagnetic interaction at this energy scale, the microscopic Amp\`ere current-current interaction.  It has time-dependent dynamical feature and can be the driving interaction of the physical, chemical or biological processes at about $1\, meV$ energy scale. A new Amp\`ere-type exchange spin interaction is also found with a magnitude about $10^{-4}$ of the well-known Heisenberg exchange spin interaction.

\end{abstract}


\maketitle


\section{Introduction} \label{sec1}

Of the four fundamental interactions \citep{Fundamentalbook}, it is the gravitational and electromagnetic interactions that govern the general physical, chemical and biological processes of the matters or the living organisms on Earth. Moreover, it is the electromagnetic interaction that dominates the physical or chemical formation of the matters as well as the biological activity processes of the living organisms. The energy scales can range from as low as about $0.1 \,  meV$ ($1 \,  K$) to as high as about $10^5  \, eV$, with examples of the former such as the biological processes in the warm-blooded animals and the latter such as the inner-shell core electrons of the Uranium atom \citep{BindingEnergyRevModPhys.39.125}. Typically, the Coulomb interaction and its various reduced ones, such as the electron-phonon interaction \citep{Mahan1990}, the Heisenberg exchange spin interaction \citep{DiracExchange}, the superexchange interaction \citep{AndersonSuperexchange}, the van der Waals interaction \citep{Langbeinbook}, {\it etc.}, are considered as the driving forces of these physical, chemical or biological processes. The spin-orbit interaction \citep{Griffithsbook} and the Breit interaction \citep{BreitPR1929,BreitPR1932} which come from the relativistic corrections are also the possible driving interactions of these processes \citep{Grant2007,JOHNSON1995255,GrzegorzHeliumPRA2001}.   

As we know that there are two fundamental macroscopic electromagnetic forces in charged systems \citep{Jackson1998}, the Coulomb force in the charge density-density channel and the Amp\`ere force in the charge current-current channel. The ratio of the Amp\`ere force to the Coulomb force is about $|\mathbf{v}_1 \cdot \mathbf{v}_2|/c^2 $, where $\mathbf{v}_{1,2}$ are the velocities of the charged particles of currents and $c$ is the speed of light. The Amp\`ere force is much smaller than the Coulomb force in the macroscopic case, where the macroscopic drift velocity $\mathbf{v}$ is much smaller than $c$. Example is such as a current $I = 1 \,  A$ in a Copper wire of $2 \,  mm$ diameter, where the macroscopic drift velocity $v \simeq 10^{-5} \, m/s \ll c$. However, in the microscopic level, the velocity of the electrons can be as high as $10^{6} \,  m/s$ in such as atomic orbitals or quantum states of metals. In this case, the ratio of the microscopic Amp\`ere force to the microscopic Coulomb force is about $10^{-4}$. As we will show in this article that the microscopic Amp\`ere and Coulomb interactions follows a similar ratio. Therefore, when the microscopic  Coulomb interaction is about $10 \,  eV$, the microscopic Amp\`ere interaction is about $1\, meV$. Thus, the microscopic Amp\`ere interaction is within the energy resolution of modern measurement techniques. It may be the driving interaction of the physical, chemical or biological processes at about $1 \,  meV$ energy scale and must be in careful consideration for these processes. 

In this article, we will revisit the fundamental microscopic Coulomb and Amp\`ere interactions in charged systems. The microscopic Coulomb interaction follows a same formula to the macroscopic case, 
\begin{equation}
V_c (t) = \frac{1}{8\pi\epsilon_0}\iint d\mathbf{r}_1 d\mathbf{r}_2 \frac{\rho(\mathbf{r}_1,t) \rho(\mathbf{r}_2, t)}{|\mathbf{r}_1-\mathbf{r}_2|} , \label{eqn1.1}
\end{equation}   
where $\rho(\mathbf{r},t)$ is the charge density. The microscopic Amp\`ere interaction has a time-dependent dynamical behavior as following:
\begin{equation}
V_a = \left(\frac{i}{\hbar}\right) \sum_{\mathbf{q}} \iint d t_1 d t_2 V_J(\mathbf{q}, t_1-t_2) \mathbf{j}_{\perp}(-\mathbf{q},t_1)\cdot \mathbf{j}_{\perp}(\mathbf{q},t_2)  , \label{eqn1.2} 
\end{equation}
where $V_J(\mathbf{q}, t) = \frac{1}{2\pi}\int d\omega V_J (\mathbf{q},\omega) e^ {-i\omega t}$ and $V_J (\mathbf{q},\omega)$ follows 
\begin{equation}
V_J (\mathbf{q},\omega) = \frac{1}{2 \epsilon_0} \, \frac{1}{\omega^2  - ( \Omega_{q}- i\delta^{+})^2 }, \,\,  \Omega_{q} = \sqrt{c^2q^2 + \omega_p^2} . \label{eqn1.3}
\end{equation}
Here $\mathbf{j}_{\perp} (\mathbf{q},t)$ is the transverse part of the charge current density and $\omega_p=\sqrt{\frac{e^2 n_e}{\epsilon_0 m}}$ is the electron plasma frequency. Physically, in the Coulomb gauge, the microscopic Coulomb interaction is a scalar potential $\phi$-field induced electromagnetic interaction in the charge density-density channel and the microscopic Amp\`ere interaction is a vector potential $\mathbf{A}$-field induced electromagnetic interaction in the charge current-current channel. The microscopic Amp\`ere current-current interaction is a dynamical manifestation of the Biot-Savart law and the Lorentz force in classical electrodynamics, where a current generates a magnetic field that acts on another current to produce a current-current Amp\`ere force. In our article, the transverse part of a vector $\mathbf{a}$ is defined by $\mathbf{a}_{\perp} \equiv \mathbf{a} \cdot (1 - \widehat{\mathbf{q}} \widehat{\mathbf{q}})$, where $\widehat{\mathbf{q}}\equiv \mathbf{q}/q$ is the unit vector of the momentum $\mathbf{q}$.    

Our article is arranged as follows. In Sec. \ref{sec2}, we will revisit the microscopic electron-electron interactions by using a path integral approach on a non-relativistic quantum electrodynamics theory, which will lead us the microscopic Coulomb and Amp\`ere interactions in the Coulomb gauge. In Sec. \ref{sec3}, we will make an estimation of the magnitude of the microscopic Amp\`ere interaction by comparison with the phonon-induced electron-electron interaction and the microscopic Coulomb interaction. In Sec. \ref{sec4}, we will derive an Amp\`ere-type exchange spin interaction, which is a new manifestation of the microscopic Amp\`ere interaction in the spin channel. Discussion on the possible roles of the Amp\`ere interaction and the further theoretical development from a relativistic quantum electrodynamics theory will be provided in Sec. \ref{sec5}, where a simple summary will also be presented.

\section{Microscopic Coulomb and Amp\`ere interactions} \label{sec2}

In this section, we will revisit the fundamental electron-electron interactions in microscopic level. In Sec. \ref{sec2.1}, we will present the microscopic action of a non-relativistic quantum electrodynamics theory. The microscopic Coulomb interaction can be easily obtained in the Coulomb gauge. In Sec. \ref{sec2.2}, we will derive the microscopic Amp\`ere interaction by using a path integral approach. In Sec. \ref{sec2.3}, we will present the microscopic Coulomb and Amp\`ere interactions for a two-component charged system with positively charged ions and negatively charged electrons.

\subsection{Microscopic action and microscopic Coulomb interaction} \label{sec2.1}

Our starting point is the charge $U(1)$ gauge invariant action $S=S_{el} +S_{em} $ for the negatively charged electrons, where
\begin{eqnarray}
S_{el} &=& \iint d\mathbf{r} dt \sum_\sigma \widebar{\Psi}_\sigma [ i\hbar \partial_t - \frac{1}{2 m} (\widehat{\mathbf{p}} + e \mathbf{A})^2 + e \phi ] \Psi_\sigma , \notag \\
S_{em} &=& \iint d\mathbf{r} dt  [ \frac{1}{2}\epsilon_0 \mathbf{E}^2 -\frac{1}{2 \mu_0} \mathbf{B}^2 ] . \label{eqn2.1} 
\end{eqnarray}
It is the microscopic action of a non-relativistic quantum electrodynamics theory with non-relativistic quantum electrons.  
Here $\Psi_\sigma=\Psi_\sigma(\mathbf{r},t)$ and $\widebar{\Psi}_\sigma=\widebar{\Psi}_\sigma(\mathbf{r},t)$ are the electron fields with spin $\sigma$, $\mathbf{A}=\mathbf{A}(\mathbf{r},t)$ and $\phi=\phi(\mathbf{r},t)$ are the respective electromagnetic vector and scalar potential fields. The electron charge $q_e = -e$ has been used. $\epsilon_0$ and $\mu_0$ are the vacuum permittivity and permeability, respectively. The electromagnetic fields $\mathbf{E}=-\boldsymbol{\nabla} \phi - \frac{\partial \mathbf{A}}{\partial t}$ and $\mathbf{B} = \boldsymbol{\nabla} \times \mathbf{A}$. It is noted that $S_{em}$ comes from the definition \citep{Altland2006} $S_{em} = \int d\mathbf{r} dt \sum_{\mu\nu} (-\frac{1}{4 \mu_0} F_{\mu \nu} F^{\mu \nu}) $, where $F_{\mu\nu}$ is the field strength tensor defined as $F_{\mu\nu}=\partial_\mu A_\nu - \partial_\nu A_\mu $ with $\mu, \nu = 0, 1, 2, 3$ and $A_\mu = (\frac{\phi}{c}, -\mathbf{A})$. Here $\partial_\nu = \frac{\partial }{\partial x^\mu}$ with $x^\mu = (ct, \mathbf{r})$. $F^{\mu\nu}$ is the contravariant form of $F_{\mu\nu}$ following $F^{\mu\nu} = \sum_{\mu^\prime \nu^\prime} \eta^{\mu\mu^\prime} \eta^{\nu \nu^\prime}  F_{\mu^\prime \nu^\prime}$, where the metric tensor $\eta^{\mu\nu} = \text{diag} (1,-1,-1,-1)$. The Euler-Lagrange equations of motion $\sum_{\mu}\partial_\mu \left[\frac{\partial \mathscr{L}}{\partial (\partial_\mu A_\nu)}\right] - \frac{\partial \mathscr{L}}{\partial A_\nu}=0$, where $\mathscr{L}$ is the Lagrangian density, lead us two Maxwell's equations, $\boldsymbol{\nabla} \cdot \mathbf{E} = \frac{\rho_e}{\epsilon_0}$ and $\boldsymbol{\nabla} \times \mathbf{B} = \mu_0 \mathbf{J}_e + \mu_0 \epsilon_0 \partial_t \mathbf{E}$. The electron charge density $\rho_e$ and charge current density $\mathbf{J}_e$ are defined by $\rho_e(\mathbf{r},t) = (-e) \sum_\sigma  \widebar{\Psi}_\sigma \Psi_\sigma$ and $\mathbf{J}_e (\mathbf{r},t) = -\frac{e}{m}\sum_\sigma  \text{Re}\, [\widebar{\Psi}_\sigma (\widehat{\mathbf{p}} + e \mathbf{A} )\Psi_\sigma ]$, where $\widehat{\mathbf{p}}=-i\hbar \boldsymbol{\nabla}$ is the momentum operator. The Bianchi identities $\partial_\mu F_{\nu\gamma} + \partial_\nu F_{\gamma\mu} + \partial_\gamma F_{\mu\nu}=0$ lead us another two Maxwell's equations, $\boldsymbol{\nabla} \times \mathbf{E} = -\frac{\partial \mathbf{B}}{\partial t}$ and $\boldsymbol{\nabla} \cdot \mathbf{B} = 0$. 

In the Coulomb gauge with $\boldsymbol{\nabla} \cdot \mathbf{A} = 0$, $\int d\mathbf{r} \frac{1}{2}\epsilon_0 \mathbf{E}^2 = \int d\mathbf{r} \frac{1}{2} \rho_e \phi + \int d\mathbf{r} \frac{1}{2}\epsilon_0 (\partial_t\mathbf{A})^2$. Since $ \boldsymbol{\nabla} \cdot \mathbf{E} =  -\boldsymbol{\nabla}^2 \phi = \frac{1}{\epsilon_0}\rho_e$, the scalar potential field is time instantaneous and follows $\phi(\mathbf{r},t) = \frac{1}{4\pi \epsilon_0} \int d\mathbf{r}^\prime \frac{\rho_e(\mathbf{r}^\prime, t)}{|\mathbf{r}-\mathbf{r}^\prime|}$. Thus the action can be reexpressed into the below form:
\begin{eqnarray}
S &=& \iint d\mathbf{r} dt\sum_{\sigma} \widebar{\Psi}_\sigma [ i\hbar \partial_t - \frac{1}{2 m} (\widehat{\mathbf{p}}+e \mathbf{A})^2 ] \Psi_\sigma - \int d t V_c(t)  \notag \\
&+& \iint d\mathbf{r} dt  [ \frac{1}{2}\epsilon_0 (\partial_t\mathbf{A})^2 -\frac{1}{2 \mu_0} (\boldsymbol{\nabla}\times \mathbf{A})^2 ] . \label{eqn2.2} 
\end{eqnarray}
Here the microscopic Coulomb interaction $V_c(t)$ is defined by Eq. (\ref{eqn1.1}) with the electron charge density $\rho_e(\mathbf{r},t)$.  

Introducing an imaginary time $\tau = i t/\hbar$, we can transform the real-time action $S$ into an imaginary-time action $S_T$ and obtain the following partition function   
\begin{equation}
e^{\frac{i}{\hbar} S } \rightarrow Z = \int  \mathscr{D} [\Psi_\sigma, \widebar{\Psi}_\sigma, \mathbf{A}] e^{-S_T[\Psi_\sigma, \widebar{\Psi}_\sigma, \mathbf{A}]} , \label{eqn2.3}
\end{equation}
where the imaginary-time action $S_T=S_1 + S_2 + S_I$ with
\begin{eqnarray}
S_1 &=&\iint d\mathbf{r} d\tau \sum_{\sigma} \widebar{\Psi}_\sigma [ \partial_\tau + \frac{\widehat{\mathbf{p}}^2}{2 m} ] \Psi_\sigma + \int d \tau V_c(\tau) ,  \notag \\
S_2 &=& \iint d\mathbf{r} d\tau   [ \frac{\epsilon_0}{2\hbar^2}(\partial_\tau \mathbf{A})^2 + \frac{1}{2 \mu_0} (\boldsymbol{\nabla}\times \mathbf{A})^2 + \frac{e^2 n_e}{2 m} \mathbf{A}^2 ] , \notag \\
S_I &=& \iint d\mathbf{r} d\tau (- \mathbf{j}_e \cdot \mathbf{A}) . \label{eqn2.4} 
\end{eqnarray}
Here $\Psi_\sigma =\Psi_\sigma(\mathbf{r},\tau)$, $\widebar{\Psi}_{\sigma}=\widebar{\Psi}_{\sigma}(\mathbf{r},\tau)$ and $\mathbf{A}=\mathbf{A}(\mathbf{r},\tau)$. To obtain $S_2$, we have made an approximation for the coupling of the electrons to the quadratic $\mathbf{A}^2$ term, where the electron density field is approximated by an average variable $n_e = \sum_\sigma \langle \widebar{\Psi}_\sigma \Psi_\sigma \rangle$. This is the only approximation we have made to obtain the microscopic Amp\`ere interaction presented in Eq. (\ref{eqn1.2}) from the non-relativistic quantum electrodynamics action in Eq. (\ref{eqn2.1}). It should be noted that the quadratic $\mathbf{A}^2$ term makes the $\mathbf{A}$ fields massive. In $S_I$, the electron charge current density $\mathbf{j}_e=\mathbf{j}_e(\mathbf{r},\tau)$ only involves the paramagnetic charge current and is defined by  
\begin{equation}
\mathbf{j}_e(\mathbf{r},\tau) = -\frac{e}{2m}\sum_\sigma [ \widebar{\Psi}_\sigma (\widehat{\mathbf{p}}\Psi_\sigma) - ( \widehat{\mathbf{p}} \widebar{\Psi}_\sigma) \Psi_\sigma ]. \label{eqn2.5}
\end{equation}

\subsection{Microscopic Amp\`ere interaction} \label{sec2.2}

Let us now consider the electromagnetic $\mathbf{A}$-field induced microscopic Amp\`ere interaction. We introduce the following Fourier transformations,
\begin{eqnarray}
\mathbf{A}(\mathbf{r},\tau) &=& \frac{1}{\sqrt{\mathcal{V}\beta}} \sum_{\mathbf{q}, i\nu_n} \mathbf{A}(\mathbf{q},i\nu_n) e^{i\mathbf{q}\cdot \mathbf{r} - i\nu_n \tau} , \notag \\
\mathbf{j}_e(\mathbf{r},\tau) &=& \frac{1}{\sqrt{\mathcal{V}\beta}} \sum_{\mathbf{q},i\nu_n} \mathbf{j}_e(\mathbf{q},i\nu_n) e^{i\mathbf{q}\cdot \mathbf{r} - i\nu_n \tau} , \label{eqn2.6} 
\end{eqnarray}
where $\mathcal{V}$ is the volume of the system and $\beta =1/k_B T$ with $T$ being the temperature. $i\nu_n$ are the bosonic imaginary Matsubara frequencies.
Now $S_2$ and $S_I$ can be reexpressed into the following forms:
\begin{eqnarray}
S_2 &=& \sum_{\mathbf{q},i\nu_n} \sum_{\alpha\beta} \widebar{\mathbf{A}}_\alpha(\mathbf{q},i\nu_n) \frac{1}{2}G_{0}^{-1}(\mathbf{q},i\nu_n) \mathbf{A}_\beta (\mathbf{q},i\nu_n)  \notag \\
&& \times \, (\delta_{\alpha\beta}-\widehat{\mathbf{q}}_\alpha \widehat{\mathbf{q}}_\beta )  \label{eqn2.7}
\end{eqnarray}
and
\begin{eqnarray}
S_I &=& -\frac{1}{2} \sum_{\mathbf{q},i\nu_n} \sum_{\alpha \beta} [\widebar{\mathbf{j}}_{e,\alpha} (\mathbf{q},i\nu_n) \mathbf{A}_\beta(\mathbf{q},i\nu_n) + \widebar{\mathbf{A}}_\alpha(\mathbf{q},i\nu_n) \notag \\ 
&& \times \, \mathbf{j}_{e,\beta}(\mathbf{q},i\nu_n)] (\delta_{\alpha\beta}-\widehat{\mathbf{q}}_\alpha \widehat{\mathbf{q}}_\beta ) . \label{eqn2.8}
\end{eqnarray}
Here $\widebar{\mathbf{A}} (\mathbf{q},i\nu_n)= \mathbf{A} (-\mathbf{q},-i\nu_n)$, $\widebar{\mathbf{j}}_e (\mathbf{q},i\nu_n)= \mathbf{j}_e (-\mathbf{q},-i\nu_n)$, and $\alpha,\beta = x,y,z$ are the indices of the vector fields. $G_0^{-1}(\mathbf{q},i\nu_n)$ is defined by
\begin{equation}
G_0^{-1}(\mathbf{q},i\nu_n) = [-\frac{\epsilon_0}{\hbar^2} (i\nu_n)^2 + \frac{\mathbf{q}^2}{\mu_0} + \frac{e^2 n_e}{m}] . \label{eqn2.9}
\end{equation}
It should be noted that here only the transverse part of the $\mathbf{A}$ fields, $\mathbf{A}_{\perp}(\mathbf{q},\tau)$, has been involved in the Coulomb gauge. 

Bilinear in the $\mathbf{A}$ fields, the Gaussian integration over the $\mathbf{A}$ fields of the partition function can now be performed straightforwardly, which yields the following expression: 
\begin{equation}
Z= C_Z \cdot \int \mathscr{D} [\Psi_\sigma, \widebar{\Psi}_\sigma] e^{-(S_1 + V_a)} , \label{eqn2.10} 
\end{equation}
where $C_Z = \prod_{\mathbf{q},i\nu_n} [\text{det}(\frac{G^{-1}}{2})]^{-1}$ and $V_a$ follows
\begin{eqnarray}
&& V_{a}  \notag \\
 &=& -\sum_{\mathbf{q},i\nu_n} \frac{1}{2}G_0(\mathbf{q},i\nu_n) [\widehat{\mathbf{q}}\times \widebar{\mathbf{j}}_e(\mathbf{q},i\nu_n)] \cdot [\widehat{\mathbf{q}} \times \mathbf{j}_e(\mathbf{q},i\nu_n))] . \notag
\end{eqnarray}
Here, the Green's function $G^{-1}$ in $C_Z$ is defined by 
\begin{equation}
G_{\alpha\beta}^{-1}(\mathbf{q},i\nu_n) = G_0^{-1}(\mathbf{q},i\nu_n)(\delta_{\alpha\beta}-\widehat{\mathbf{q}}_\alpha \widehat{\mathbf{q}}_\beta ) . \label{eqn2.11}
\end{equation}
It is noted that $C_Z$ describes the contribution of the $\mathbf{A}$-field quantum fluctuations to the partition function. Thus, there is an $\mathbf{A}$-field induced microscopic Amp\`ere interaction in the current-current channel 
\begin{equation}
V_{a} = \sum_{\mathbf{q},i\nu_n} V_{J}(\mathbf{q},i\nu_n) [\widehat{\mathbf{q}}\times \widebar{\mathbf{j}}_e(\mathbf{q},i\nu_n)] \cdot [\widehat{\mathbf{q}} \times \mathbf{j}_e(\mathbf{q},i\nu_n))] , \label{eqn2.12}
\end{equation}
where $V_{J}(\mathbf{q},i\nu_n))$ is defined as
\begin{equation}
V_{J}(\mathbf{q},i\nu_n)) = \frac{\hbar^2}{2\epsilon_0} \frac{1}{(i\nu_n)^2 - [(\hbar c q)^2 + (\hbar \omega_p)^2 ] } . \label{eqn2.13}
\end{equation}
Since $(\widehat{\mathbf{q}}\times \mathbf{j}_1) \cdot (\widehat{\mathbf{q}} \times \mathbf{j}_2) = \sum_{\alpha\beta} \mathbf{j}_{1 \alpha} \mathbf{j}_{2 \beta} (\delta_{\alpha\beta}-\widehat{\mathbf{q}}_\alpha \widehat{\mathbf{q}}_\beta ) =  \mathbf{j}_{1 \perp} \cdot \mathbf{j}_{2 \perp}$, $V_a$ has another form:
\begin{equation}
V_{a} = \sum_{\mathbf{q},i\nu_n} V_{J}(\mathbf{q},i\nu_n) \, \widebar{\mathbf{j}}_{e\perp}(\mathbf{q}, i\nu_n) \cdot \mathbf{j}_{e \perp}(\mathbf{q},i\nu_n) . \label{eqn2.14}
\end{equation}
Thus, the microscopic Amp\`ere interaction only involves the transverse part of the charge current density in the Coulomb gauge. 

In the static case with $i\nu_n = 0$, we can show from Eq. (\ref{eqn2.13}) that
\begin{equation}
V_{J}(\mathbf{r}, 0)\equiv \frac{1}{\mathcal{V}}\sum_{\mathbf{q}} V_{J}(\mathbf{q},0) e^{i\mathbf{q}\cdot \mathbf{r}}=-\frac{\mu_0}{8\pi}\frac{1}{r} e^{-q_F r} . \label{eqn2.15}
\end{equation}
Here the exponential decay factor $e^{-q_F r}$ with $q_F= \omega_p/c$ stems from the screening effects of the electron charge density fluctuations. The static microscopic Amp\`ere current-current interaction can be shown to follow 
\begin{equation}
V_{a} =-\frac{\mu_0}{8\pi} \iint d\mathbf{r}_1 d \mathbf{r}_2  \, \frac{1}{r} e^{-q_F r}\, \mathbf{j}_{e\perp} (\mathbf{r}_1,0) \cdot \mathbf{j}_{e\perp} (\mathbf{r}_2,0) , \label{eqn2.16}
\end{equation}
where $r =|\mathbf{r}_1 - \mathbf{r}_2|$ and $\mathbf{j}_{e\perp} (\mathbf{r},0)$ is defined as $\mathbf{j}_{e\perp} (\mathbf{r},0) = \frac{1}{\sqrt{\mathcal{V}}}\sum_{\mathbf{q}} \mathbf{j}_{e\perp} (\mathbf{q},0) e^{i\mathbf{q}\cdot \mathbf{r}}$. 

When we introduce the transformations 
\begin{eqnarray}
&&\mathbf{j}_{e\perp}(\mathbf{q},i\nu_n) = \frac{1}{\sqrt{\beta}} \int_0^{\beta} d\tau \, \mathbf{j}_{e\perp}(\mathbf{q},\tau) e^{i\nu_n \tau} , \notag \\
&&V_{J}(\mathbf{q},\tau) = \frac{1}{\beta} \sum_{i\nu_n} V_{J}(\mathbf{q},i\nu_n) e^{-i\nu_n \tau} , \label{eqn2.17}
\end{eqnarray}
the microscopic Amp\`ere interaction $V_a$ of Eq. (\ref{eqn2.14}) can be transformed into the following imaginary-time form: 
\begin{equation}
V_{a} = \sum_{\mathbf{q}}\iint d\tau_1 d\tau_2 V_{J}(\mathbf{q},\tau_1-\tau_2) \, \widebar{\mathbf{j}}_{e\perp}(\mathbf{q}, \tau_1) \cdot \mathbf{j}_{e \perp}(\mathbf{q},\tau_2) . \label{eqn2.18}
\end{equation}
Making the imaginary times back into the real ones by $\tau=i t/\hbar$, we can obtain the real-time microscopic Amp\`ere interaction presented in Eq. (\ref{eqn1.2}), where the electron charge current density $\mathbf{j}_{e}(\mathbf{q},t) = \frac{1}{\sqrt{\mathcal{V}}} \int d\mathbf{r} \, \mathbf{j}_{e}(\mathbf{r},t) e^{-i\mathbf{q}\cdot \mathbf{r}}$. Here $\mathbf{j}_{e}(\mathbf{r},t)$ is defined similarly to $\mathbf{j}_{e}(\mathbf{r},\tau)$ of Eq. (\ref{eqn2.5}).  

For the static case with only contribution from $i\nu_n =0$, $\mathbf{j}_{e\perp}(\mathbf{r},\tau)=\frac{1}{\sqrt{\beta}} \mathbf{j}_{e\perp}(\mathbf{r},0)$ which is time independent. By considering the transformation $\beta=\int d\tau \rightarrow \frac{i}{\hbar} \int d t$ and making the definition $\mathbf{j}_{e}(\mathbf{r})\equiv \mathbf{j}_{e\perp}(\mathbf{r},t)$, we can obtain the real-time microscopic Amp\`ere interaction in the static case from Eq. (\ref{eqn2.16}) as following: $V_a = \frac{i}{\hbar}\int d t V_{a,s}$, where 
\begin{equation}
V_{a,s} = -\frac{\mu_0}{8\pi} \iint d\mathbf{r}_1 d \mathbf{r}_2  \, \frac{1}{r} e^{-q_F r}\, \mathbf{j}_{e} (\mathbf{r}_1) \cdot \mathbf{j}_{e} (\mathbf{r}_2) . \label{eqn2.19}
\end{equation}
It shows that two parallel currents with a same direction have attractive interaction, a well-known result in classical electrodynamics \citep{Jackson1998}. It should be noted that the steady current density in the static case is always transverse due to the local charge conservation law $\frac{\partial \rho_e}{\partial t} + \boldsymbol{\nabla} \cdot \mathbf{j}_{e} = 0$ which leads to $\boldsymbol{\nabla} \cdot \mathbf{j}_{e} = 0$. 

At the end of this subsection, some remarks are made as follows. First, the microscopic Coulomb interaction $V_c$ and the microscopic Amp\`ere interaction $V_a$ we have obtained are gauge independent. Consider an arbitrary gauge transformation with $\Psi_\sigma(\mathbf{r},t) \rightarrow \Psi^\prime_\sigma(\mathbf{r},t) = \Psi_\sigma(\mathbf{r},t) e^{i \theta(\mathbf{r},t)}$ and $\widebar{\Psi}_\sigma(\mathbf{r},t) \rightarrow \widebar{\Psi}^\prime_\sigma(\mathbf{r},t) = \widebar{\Psi}_\sigma(\mathbf{r},t) e^{-i\theta(\mathbf{r}, t)}$. It can be easily shown that the electron charge density $\rho_e (\mathbf{r},t)$ and the transverse part of the electron charge current density $\mathbf{j}_{e\perp}(\mathbf{r},t)$ are gauge invariant. Since $V_c$ is a function of $\rho_e(\mathbf{r},t)$ and $V_a$ is a function of $\mathbf{j}_{e\perp}(\mathbf{r},t)$, they are also gauge invariant under the arbitrary gauge transformation. The same conclusion can be made for the imaginary time results. Second, the microscopic Coulomb interaction $V_c$ can be obtained by a path integral over the scalar $\phi$ fields. In this case, there is a $\phi$-field relevant contribution to the partition function $Z_\phi = \int \mathscr{D} [\phi] \exp (- S_\phi)$, where the action $S_\phi = \iint d\mathbf{r} d\tau [\rho_e \phi + \frac{1}{2} \epsilon_0 \phi (\boldsymbol{\nabla}^2 \phi)]$. With a similar procedure for the path integral over the $\mathbf{A}$ fields, we can show that $Z_\phi = C_\phi \exp (-\int d\tau V_c (\tau))$, where $C_\phi = \prod_{\mathbf{q},i\nu_n} \left(-\frac{1}{2} \epsilon_0 q^2\right)^{-1}$ describes the $\phi$-field quantum fluctuation contribution to the partition function. Therefore, the path integral over of the $\phi$ fields can yield the $\phi$-field induced microscopic Coulomb interaction $V_c$. Third, the microscopic electromagnetic interactions we have obtained are consistent with the results obtained by Feynman \citep{Feynman1950, FeynmanBook2010}. The microscopic Amp\`ere interaction in Eq. (\ref{eqn1.2}) follows $\exp (-A_a) = \exp (\frac{i}{\hbar} I_a)$, where the effective current-current interaction $I_a$ is defined by  
\begin{eqnarray}
I_a &=& \sum_{\mathbf{q}} \iint d t_1 d t_2 \frac{i}{4\epsilon_0 \Omega_q} \mathbf{j}_{e\perp} (-\mathbf{q},t_1)\cdot \mathbf{j}_{e\perp} (\mathbf{q},t_2) \notag \\
&& \times \, e^{-(i \Omega_q + \delta^+) |t_1 - t_2|} . \label{eqn2.20}
\end{eqnarray}     
When the mass of the $\mathbf{A}$ fields is ignored by setting $\omega_p=0$, $I_a$ recovers the result obtained by Feynman (with a different factor due to the difference in the SI and Gaussian units), the latter of which is obtained from a classical action by a real-time path integral approach \citep{FeynmanBook2010}.

\subsection{Two-component charged system} \label{sec2.3}

In the above Sec. \ref{sec2.1} and Sec. \ref{sec2.2}, we have presented the microscopic Coulomb and Amp\`ere interactions for the negatively charged electrons, each of which has charge $q_e = -e$. We will now present a simple extension to a two-component charged system of positively charged ions and negatively charged electrons, each of the formers has charge $q_I = + Z e$.   

In this two-component charged system, there are additional couplings of the positively charged ions to the electromagnetic fields $\phi$ and $\mathbf{A}$: $\rho_I \phi$, $-\mathbf{j}_{I}\cdot \mathbf{A}$ and $\frac{q_I^2 n_I }{2 M} \mathbf{A}^2$. Here $\rho_I$ is the ion charge density, $\mathbf{j}_{I}$ is the ion charge current density, $n_I$ is the ion density and $M$ is the ion mass. A similar derivation leads us the same expressions of the microscopic Coulomb interaction in Eq. (\ref{eqn1.1}) and the microscopic Amp\`ere interaction in Eqs. (\ref{eqn1.2}), (\ref{eqn2.14}) and (\ref{eqn2.18}) with the following redefinitions:
\begin{equation}
\rho = \rho_e + \rho_I , \ \,  \mathbf{j} = \mathbf{j}_e + \mathbf{j}_I , \label{eqn2.20}
\end{equation}
and $\omega_p^2$ in $V_J(\mathbf{q}, \omega)$ of Eq. (\ref{eqn1.3}) and $V_J(\mathbf{q}, i\nu_n)$ of Eq. (\ref{eqn2.13}) are modified into $\omega_p ^2 + \omega_I^2$ with the ion plasma frequency defined by $\omega_I = \sqrt{\frac{q_I^2 n_I}{\epsilon_0 M}}$. 

In most cases of condensed matters, because $m\ll M$ and $\frac{Z^2 n_I m}{n_e M} \ll 1$, we have $\omega_I^2 \ll \omega_p^2$. In these cases, we can ignore $\omega_I^2$ in $V_J(\mathbf{q}, \omega)$ and $V_J(\mathbf{q}, i\nu_n)$. Moreover, as $m\ll M$ also leads the velocity of the ions $v_I$ to be much smaller than the velocity of the electrons $v_e$, the magnitude of the ion charge current density $\mathbf{j}_I = \rho_I \mathbf{v}_I $ is also much smaller than that of the electron charge current density $\mathbf{j}_e = \rho_e \mathbf{v}_e$. Therefore, the contribution of the ion charge current to the microscopic Amp\`ere interaction can be neglected in condensed matters. This leads us the simplified expression of $V_a$ in Eqs. (\ref{eqn1.2}) and (\ref{eqn1.3}) for condensed matters.

\section{Estimation of microscopic Amp\`ere interaction} \label{sec3}

We will present an estimation of the magnitude of the microscopic Amp\`ere interaction in this section. The phonon-induced electron-electron interaction and the microscopic Coulomb interaction will be considered as comparative references.  

\subsection{Comparison with phonon-induced electron-electron interaction} \label{sec3.1}

Let us consider the electrons in crystal condensed matters. For the electrons in the plane-wave states, $\Psi_\sigma(\mathbf{r},\tau) = \frac{1}{\sqrt{\mathcal{V}}}\sum_{\mathbf{k}} e^{i\mathbf{k}\cdot\mathbf{r}} c_{\mathbf{k}\sigma}$, where $c_{\mathbf{k}\sigma}$ is the annihilation operator for the electrons with momentum $\mathbf{k}$ and spin $\sigma$. The electron charge current density $\mathbf{j}_e(\mathbf{q},\tau)$ follows
\begin{equation}
\mathbf{j}_e(\mathbf{q},\tau) = -\frac{e \hbar}{m \sqrt{\mathcal{V}}}\sum_{\mathbf{k}\sigma} ( \mathbf{k}+\frac{\mathbf{q}}{2}) c_{\mathbf{k}\sigma}^{\dag}(\tau) c_{\mathbf{k}+\mathbf{q} \sigma} (\tau) . \label{eqn3.1}
\end{equation}
The microscopic Amp\`ere interaction $V_{a}$ in Eq. (\ref{eqn2.18}) can be expressed into the following second-quantization form:
\begin{eqnarray}
V_{a} &=& \sum_{\mathbf{k}_{1}\mathbf{k}_2\mathbf{q}\sigma_1\sigma_2}\iint d \tau_1 d\tau_2 \ g_J(\mathbf{k}_1,\mathbf{k}_2,\mathbf{q};\tau_1,\tau_2) \notag \\
&\times & c_{\mathbf{k}_1\sigma_1}^{\dag}(\tau_1) c_{\mathbf{k}_2\sigma_2}^{\dag}(\tau_2) c_{\mathbf{k}_2 + \mathbf{q}\sigma_2}(\tau_2) c_{\mathbf{k}_1 - \mathbf{q}\sigma_1}(\tau_1) ,   \label{eqn3.2}
\end{eqnarray}
where the interaction matrix elements $g_J(\mathbf{k}_1,\mathbf{k}_2,\mathbf{q};\tau_1,\tau_2)$ are defined as 
\begin{equation}
g_J = \frac{e^2\hbar^2}{\mathcal{V} m^2} V_J(\mathbf{q},\tau_1-\tau_2) [ (\mathbf{k}_1 -\frac{1}{2}\mathbf{\mathbf{q}})_{\perp} \cdot (\mathbf{k}_2 +\frac{1}{2}\mathbf{q})_{\perp} ] . \label{eqn3.3} 
\end{equation}
It should be noted that for the general Bloch-band electrons, the electron charge current density follows 
\begin{equation}
\mathbf{j}_e(\mathbf{r},\tau) = -\frac{e}{2 \sqrt{\mathcal{V}}}\sum_{\mathbf{k}\sigma} (\mathbf{v}_\mathbf{k} + \mathbf{v}_{\mathbf{k}+\mathbf{q}}) c_{\mathbf{k}\sigma}^{\dag} (\tau) c_{\mathbf{k}+\mathbf{q} \sigma} (\tau) , \label{eqn3.4}
\end{equation}
where the velocity of the Bloch-band electrons is defined by $\mathbf{v}_{\mathbf{k}}\equiv \frac{1}{\hbar}\boldsymbol{\nabla}_{\mathbf{k}} \varepsilon_{\mathbf{k}}$. In this case, the interaction matrix elements $g_J$ of $V_a$ are defined by 
\begin{equation}
g_J = \frac{e^2}{4\mathcal{V}} V_{J}(\mathbf{q},\tau_1-\tau_2) [(\mathbf{v}_{\mathbf{k}_1}+\mathbf{v}_{\mathbf{k}_1-\mathbf{q}})_{\perp}\cdot (\mathbf{v}_{\mathbf{k}_2}+\mathbf{v}_{\mathbf{k}_2+\mathbf{q}})_{\perp} ] .  \label{eqn3.5} 
\end{equation}

Now let us consider the electron-lattice-ion interaction $V(\mathbf{r})=\sum_{j} V_L(\mathbf{r}-\mathbf{r}_j)$, where $V_L(\mathbf{r}-\mathbf{r}_j)$ is the interaction of the $j$-th ion at position $\mathbf{r}_j$ upon the electron at position $\mathbf{r}$. Introduce the Fourier transformation $V_L(\mathbf{q})=\frac{1}{\mathcal{V}}\int d\mathbf{r} \sum_j V_L(\mathbf{r}-\mathbf{r}_j) e^{-i\mathbf{q}\cdot (\mathbf{r}-\mathbf{r}_j)}$ and reexpress $\mathbf{r}_j$ as $\mathbf{r}_j=\mathbf{R}_j + \mathbf{u}_j$ where $\mathbf{R}_j$ is the crystal lattice vector and $\mathbf{u}_j$ is the ion displacement. Following a standard treatment ({\it e.g.}, see references \citep{Mahan1990} and \citep{Colemanbook0}), we can expand $V_L(\mathbf{r}-\mathbf{r}_j)$ to linear terms of $\mathbf{u}_j$ and quantize these terms by introducing the electron and phonon field operators. Then, we can obtain the electron-phonon interaction as 
\begin{equation}
V_{ep} = -\sum_{\mathbf{k}\mathbf{q}\lambda\sigma} \frac{i g_{\lambda}(\mathbf{q})}{\sqrt{N}}  c^{\dag}_{\mathbf{k+q}\sigma} (\tau) c_{\mathbf{k}} (\tau) [b_{\mathbf{q}\lambda}(\tau) + b^\dag_{-\mathbf{q}\lambda}(\tau)] , \label{eqn3.6} 
\end{equation}
where $g_{\lambda}(\mathbf{q})=[\frac{\hbar}{ M  \omega_{\lambda}(\mathbf{q})}]^{1/2} [\mathbf{q}\cdot\boldsymbol{\epsilon}_\lambda(\mathbf{q})] V_{L}(\mathbf{q})$, $N$ is the lattice-ion number, $b$ and $b^\dag$ are the annihilation and creation operators of phonons. In $g_{\lambda}(\mathbf{q})$, $M$ is the ion mass, $\omega_\lambda(\mathbf{q})$ is the frequency dispersion of the $\lambda$-th branch phonons and $\boldsymbol{\epsilon}_\lambda$ is the corresponding polarization vector. Here we assume one ion in each unit cell.

Consider the second-order perturbations of the electron-phonon interaction $V_{ep}$, we can obtain an effective phonon-induced electron-electron interaction $V_{p}$, 
\begin{eqnarray}
V_{p} &=& \sum_{\mathbf{k}_{1}\mathbf{k}_2\mathbf{q}\sigma_1\sigma_2}\int d \tau_1 d\tau_2 \ g_{p}(\mathbf{q};\tau_1,\tau_2) \notag \\
& \times & c_{\mathbf{k}_1\sigma_1}^{\dag}(\tau_1) c_{\mathbf{k}_2\sigma_2}^{\dag}(\tau_2) c_{\mathbf{k}_2 + \mathbf{q}\sigma_2}(\tau_2) c_{\mathbf{k}_1 - \mathbf{q}\sigma_1}(\tau_1) ,   \label{eqn3.7}
\end{eqnarray}
where the interaction matrix elements $g_{p}(\mathbf{q};\tau_1,\tau_2)$ are defined by 
\begin{eqnarray}
g_p (\mathbf{q};\tau_1,\tau_2) = \sum_{i\nu_n \lambda} \frac{\hbar^2 \vert \mathbf{q}\cdot\boldsymbol{\epsilon}_\lambda(\mathbf{q})\vert^2 [V_{L}(\mathbf{q})]^2}{\beta N M [(i\nu_n)^2 -\hbar^2 \omega^2_{\lambda}(\mathbf{q})]} e^{-i\nu_n (\tau_1 -\tau_2)}.  \notag \\
 \label{eqn3.8} 
\end{eqnarray}

Let us consider the ratio of the microscopic Amp\`ere interaction $V_a$ to the phonon-induced electron-electron interaction $V_{p}$ for the case with $i\nu_n \rightarrow 0$. Introduce the following approximate interaction constants $\widebar{g}_{J}$ and $\widebar{g}_{p}$ for $g_J$ and $g_p$, respectively, 
\begin{eqnarray}
&& \widebar{g}_{J} = \frac{e^2 \hbar^2 k_F^2}{\beta\mathcal{V} m^2}\frac{1}{ 2\epsilon_0 \omega_p^2} , \notag \\
&&  \widebar{g}_{p} = \frac{1}{\beta N  \langle\hbar\omega_{\lambda}(\mathbf{q})\rangle} \langle g_{\lambda}(\mathbf{q})\rangle^2 . \label{eqn3.9}
\end{eqnarray}
Here we have made approximation that $\mathbf{k}_1$ and $\mathbf{k}_2$ are approximated by the Fermi momentum $k_F$ and $\mathbf{q}=0$ in $g_{J}$. We introduce $\langle \hbar\omega_\lambda(\mathbf{q})\rangle$ and $\langle g_{\lambda}(\mathbf{q})\rangle$ to estimate $g_p$. Here $\langle x_\lambda(\mathbf{q}) \rangle$ represents the phonon momentum $\mathbf{q}$ and polarization $\lambda$ average. The ratio of the magnitude of $V_a$ to that of $V_p$ can be estimated approximately by 
\begin{eqnarray}
\alpha_1 = \frac{\widebar{g}_J}{\widebar{g}_p} = \frac{E_F\cdot \langle \hbar \omega_{\lambda}(\mathbf{q})\rangle}{ \langle g_{\lambda}(\mathbf{q})\rangle^2} , \label{eqn3.10}
\end{eqnarray}
where $E_F=\hbar^2 k_F^2/2 m$ is the Fermi energy. 

From the references \cite{LeePickett2004,BlaseAdessi2004,GiustinoCohenLouie2007}, we can approximate the electron-phonon interaction to be in the range $\langle g_{\lambda}(\mathbf{q})\rangle \in (0.1eV,1eV)$. The phonon energy $\langle \hbar \omega_{\lambda}(\mathbf{q})\rangle$ can be approximated in magnitude by
\begin{equation}
\langle \hbar \omega_{\lambda}(\mathbf{q})\rangle \simeq \hbar \omega_D = k_B \theta_D , \label{eqn3.11}
\end{equation}
where $\omega_D$ is the Debye frequency and $\theta_D$ is the Debye temperature. As an example, let us consider the metallic Sn, whose Debye temperature $\theta_D = 200K$ \citep{WEBB201517} and the Fermi energy $E_F = 10.2eV$ \citep{AshcroftMermin1976,FermiEnergyFromAshcroftMermin1976}.  The ratio $\alpha_1$ for the metallic Sn can be estimated following Eq. (\ref{eqn3.10}):
\begin{equation}
\alpha_1 = \left\{
\begin{array}{l}
17.60, \,  \ \text{when} \ \langle g_{\lambda}(\mathbf{q})\rangle = 0.1 eV , \\
0.70, \quad \text{when} \ \langle g_{\lambda}(\mathbf{q})\rangle = 0.5 eV , \\
0.18, \quad \text{when} \ \langle g_{\lambda}(\mathbf{q})\rangle = 1.0 eV . \\
\end{array} 
\right. \label{eqn3.12}
\end{equation} 
This result shows that the microscopic Amp\`ere interaction induced by the fluctuations of the electromagnetic $\mathbf{A}$ fields in the low-frequency long-wavelength limit has a magnitude of order about the phonon-induced electron-electron interaction.  

\subsection{Comparison with microscopic Coulomb interaction}

Consider two point-like electrons with velocities $\mathbf{v}_1$ and $\mathbf{v}_2$, respectively. The electron charge current density is now defined by $\mathbf{j}_e(\mathbf{r}) = -e \delta (\mathbf{r}-\mathbf{r}_1(t)) \mathbf{v}_1 - e\delta (\mathbf{r}-\mathbf{r}_2(t)) \mathbf{v}_2$. From Eq. (\ref{eqn2.19}), the microscopic Amp\`ere interaction of two moving electrons in the static case follows
\begin{equation}
V_{a} = - \frac{\mu_0}{4\pi} \mathbf{v}_1 \cdot \mathbf{v}_2 \frac{e^2}{r} ,   \label{eqn3.13}
\end{equation}
where $r = \vert \mathbf{r}_1 - \mathbf{r}_2 \vert$. 
Since the two electron microscopic Coulomb interaction follows
\begin{equation}
V_c = \frac{1}{4\pi \epsilon_0} \frac{e^2}{r} , \label{eqn3.14}
\end{equation}
the ratio of the microscopic Amp\`ere interaction to the microscopic Coulomb interaction of two electrons in the static case follows 
\begin{equation}
\alpha_2 = \frac{|V_{a}|}{|V_c|} = \frac{| \mathbf{v}_1 \cdot \mathbf{v}_2 |}{c^2} . \label{eqn3.15}
\end{equation}

For the electrons in metals, $m v_F \simeq \hbar k_F$ where $v_F$ is the Fermi velocity. $v_F \simeq 10^{6} \, m/s$ \citep{AshcroftMermin1976,FermiEnergyFromAshcroftMermin1976} as $k_F \simeq 1/a$ and the lattice constant $a\simeq 10^{-10} \, m$. In these cases, $E_F=\hbar^2 k_F^2/2 m \simeq 10 \, eV$. As the average microscopic Coulomb interaction of two electrons has a similar magnitude to the electron-ion interaction, we can use $E_F$ to estimate $V_c$. When $v_1\simeq v_2 \simeq v_F$, the microscopic Amp\`ere interaction of two electrons near the Fermi energy in metals follows $V_{a}\simeq 10^{-4}  V_c \simeq 1 \, meV$. 

For the electrons in light hydrogen-like atomic orbital states with angular momentum $l\hbar = \vert \mathbf{r} \times m \mathbf{v} \vert $, $v \simeq 10^{6} \, m/s$ for $l\simeq 1$ and $r \simeq 10^{-10} \, m$. The Coulomb interaction can be approximated by the eigenvalues of the atomic orbital states, thus $V_c \simeq 10 \, eV$. Therefore, the microscopic Amp\`ere interaction of two electrons in light hydrogen-like atomic orbital states has a magnitude at the order of $1 \, meV$. As the ratio of the microscopic Amp\`ere interaction to the microscopic Coulomb interaction $\alpha_2=|\mathbf{v}_1 \cdot \mathbf{v}_2 |/c^2$, the microscopic Amp\`ere interaction has a magnitude at the order of the $(v/c)^2$ relativistic corrections.    
As previously shown \citep{Grant2007,JOHNSON1995255,GrzegorzHeliumPRA2001}, the atoms or molecules containing heavier elements such as transition metals, lanthanides or actinides have large relativistic corrections in their energy spectra. Thus, the microscopic Amp\`ere interaction will play important roles in these atoms or molecules with higher magnitude of nuclear charge $Z$.

\section{Amp\`ere-type exchange spin interaction}\label{sec4}

Let us define the expansion of the electron fields by $\Psi_\sigma(\mathbf{r},t) = \sum_{l} \psi_l(\mathbf{r}) c_{l \sigma}(t)$ and $\widebar{\Psi}_\sigma(\mathbf{r},t) = \sum_{l} \psi^{\ast}_l(\mathbf{r}) c^\dag_{l \sigma}(t)$, where $\psi_l(\mathbf{r})$ are basis wave functions. $\psi_l(\mathbf{r})$ can be the local Wannier wave functions in condensed matters or the single-electron eigenfunctions in atoms. The microscopic Coulomb interaction $V_c$ can be expressed into the following form: 
\begin{eqnarray}
V_c &=& \frac{1}{8\pi \epsilon_0} \sum_{l_i \sigma_1 \sigma_2} \iint d\mathbf{r}_1 d\mathbf{r}_2 \frac{e^2}{r} \psi^{\ast}_{l_1}(\mathbf{r}_1) \psi_{l_2}(\mathbf{r}_1) \psi^{\ast}_{l_3}(\mathbf{r}_2) \psi_{l_4}(\mathbf{r}_2)  \notag \\
&& \times :c^\dag_{l_1 \sigma_1} c_{l_2 \sigma_1} c^\dag_{l_3 \sigma_2} c_{l_4 \sigma_2}: ,   \label{eqn4.1}
\end{eqnarray}   
where $:\, :$ represents the normal ordering operation.
In the channels with $l_1 = l_2$ and $l_3=l_4$, we can obtain the Coulomb-type Hubbard-like interaction 
\begin{equation}
V_{c,1} = \frac{1}{2}\sum_{l_1 \not= l_2} U^{c}_{l_1 l_2} \, n_{l_1} n_{l_2} + \sum_l U^c_{ll} \, n_{l\uparrow} n_{l\downarrow} , \label{eqn4.2}
\end{equation} 
where $U^{c}_{l_1 l_2}$ is defined by 
\begin{equation}
U^{c}_{l_1 l_2} = \frac{1}{4\pi \epsilon_0} \iint d\mathbf{r}_1 d\mathbf{r}_2 \frac{e^2}{r} |\psi_{l_1}(\mathbf{r}_1)|^2 |\psi_{l_2}(\mathbf{r}_2)|^2 , \label{eqn4.3} 
\end{equation}
and $n_l = \sum_{\sigma} c^\dag_{l \sigma} c_{l \sigma}$. In the channels with $l_1=l_4$ and $l_2=l_3$, we can obtain the Coulomb-type Heisenberg exchange spin interaction
\begin{equation}
V_{c,2} = - \sum_{l_1\not= l_2} J^{c}_{l_1 l_2} \, (\mathbf{S}_{l_1} \cdot \mathbf{S}_{l_2} + \frac{1}{4} n_{l_1} n_{l_2}) + \sum_l J^c_{ll} \, (S^z_{l})^2  , \label{eqn4.4}
\end{equation}
where $J^{c}_{l_1 l_2}$ is defined by 
\begin{equation}
J^{c}_{l_1 l_2} = \frac{1}{4\pi \epsilon_0} \iint d\mathbf{r}_1 d\mathbf{r}_2 \frac{e^2}{r} \psi^{\ast}_{l_1}(\mathbf{r}_1) \psi_{l_2}(\mathbf{r}_1) \psi^{\ast}_{l_2}(\mathbf{r}_2) \psi_{l_1}(\mathbf{r}_2) , \label{eqn4.5} 
\end{equation}
and the spin operators $\mathbf{S}_{l}$ is defined by $\mathbf{S}_{l} = \frac{1}{2} \sum_{\sigma_1\sigma_2} c^\dag_{l\sigma_1} \boldsymbol{\tau}_{\sigma_1\sigma_2} c_{l \sigma_2}$ with $\boldsymbol{\tau}$ being the Pauli matrices. In Eq. (\ref{eqn4.4}), a constant contribution $\alpha_c= -\frac{3}{4}\sum_l J^c_{ll}$ to $V_{c,2}$ is ignored. 

A similar consideration can be made on the static microscopic Amp\`ere interaction by using the above procedure. Eq. (\ref{eqn2.19}) has the following expansion expression:
\begin{eqnarray}
V_{a,s} &=& -\frac{\mu_0}{8\pi} \sum_{l_i \sigma_1 \sigma_2} \iint d\mathbf{r}_1 d\mathbf{r}_2 \frac{e^2}{4 m^2 r} e^{-q_F r} \widehat{\mathbf{p}}_{l_1 l_2}(\mathbf{r}_1) \cdot \widehat{\mathbf{p}}_{l_3 l_4}(\mathbf{r}_2)\notag \\
&& \times :c^\dag_{l_1 \sigma_1} c_{l_2 \sigma_1} c^\dag_{l_3 \sigma_2} c_{l_4 \sigma_2}: ,   \label{eqn4.6}
\end{eqnarray}
where $\widehat{\mathbf{p}}_{l l^\prime}(\mathbf{r})$ is defined by 
\begin{equation}
\widehat{\mathbf{p}}_{l l^\prime}(\mathbf{r}) = \psi^{\ast}_{l}(\mathbf{r}) [\widehat{\mathbf{p}} \psi_{l^\prime}(\mathbf{r})] - [\widehat{\mathbf{p}} \psi^{\ast}_{l}(\mathbf{r}) ] \psi_{l^\prime}(\mathbf{r}) . \label{eqn4.7}
\end{equation} 
In the channels with $l_1 = l_2$ and $l_3=l_4$, we can obtain an Amp\`ere-type Hubbard-like interaction
\begin{equation}
V_{a,1} = - \frac{1}{2}\sum_{l_1\not= l_2} U^{a}_{l_1 l_2} \, n_{l_1} n_{l_2} -\sum_l U^a_{ll} \, n_{l\uparrow} n_{l\downarrow} , \label{eqn4.8}
\end{equation} 
where $U^{a}_{l_1 l_2}$ is defined by 
\begin{equation}
U^{a}_{l_1 l_2} = \frac{\mu_0}{4\pi} \iint d\mathbf{r}_1 d\mathbf{r}_2 \frac{e^2}{4 m^2 r} e^{-q_F r} \widehat{\mathbf{p}}_{l_1 l_1}(\mathbf{r}_1) \cdot \widehat{\mathbf{p}}_{l_2 l_2}(\mathbf{r}_2) . \label{eqn4.9} 
\end{equation} 
Similarly, in the channels with $l_1=l_4$ and $l_2=l_3$, we can obtain an Amp\`ere-type exchange spin interaction
\begin{equation}
V_{a,2} = \sum_{l_1\not= l_2} J^{a}_{l_1 l_2} \, (\mathbf{S}_{l_1} \cdot \mathbf{S}_{l_2} + \frac{1}{4} n_{l_1} n_{l_2}) -\sum_{l} J^a_{ll} \, (S^z_l)^2 , \label{eqn4.10}
\end{equation}
where $J^{a}_{l_1 l_2}$ is defined by 
\begin{equation}
J^{a}_{l_1 l_2} = \frac{\mu_0}{4\pi} \iint d\mathbf{r}_1 d\mathbf{r}_2 \frac{e^2}{4 m^2 r} e^{-q_F r} \widehat{\mathbf{p}}_{l_1 l_2}(\mathbf{r}_1) \cdot \widehat{\mathbf{p}}_{l_2 l_1}(\mathbf{r}_2) . \label{eqn4.11} 
\end{equation}
Similarly, a constant contribution $\alpha_a = \frac{3}{4}\sum_l J^a_{ll}$ to $V_{a,2}$ in Eq. (\ref{eqn4.10}) is ignored.  

From the rough estimations by using the procedures presented in Sec. \ref{sec3}, we have $|U^a|/|U^c|, |J^a|/|J^c|\simeq |\mathbf{v}_1\cdot \mathbf{v}_2|/c^2$, where the two velocities are relevant to the electron charge current densities $\mathbf{j}_{e}= \rho_{e} \mathbf{v}$. In the cases with $v_1\simeq v_2 \simeq 10^6 \, m/s$, $U^a\simeq 10^{-4} U^c$ and $J^a\simeq 10^{-4} J^c$. When $U^c\simeq \, 10 eV$, $U^a \simeq \, 1 meV$. When $J^c \simeq \, 100 meV$, $J^a \simeq \, 0.01 meV$ which is a very small energy scale.

\section{Discussion and Summary} \label{sec5} 

In the Coulomb gauge, the microscopic Coulomb interaction is instantaneous and the microscopic Amp\`ere interaction is time-retarded dynamical. Physically, the time-retarded dynamics of the microscopic Amp\`ere interaction stems from the propagating of the photons, the quanta of the electromagnetic fields, which become massive in condensed matters due to the coupling to the charged particles. 

Based on the phonon-induced electron-electron interaction in the pairing channel \citep{BardeenPines1955}, an effective pairing interaction for the Cooper-pair superconductivity can be obtained as 
\begin{equation}
V_{1} = -\frac{1}{N}\sum_{\mathbf{k}\mathbf{k}^\prime} g^{(0)}_{p} c^\dag_{\mathbf{k}\uparrow} c^{\dag}_{-\mathbf{k}\downarrow} c_{-\mathbf{k}^\prime \downarrow} c_{\mathbf{k}^\prime \uparrow} , \label{eqn5.1} 
\end{equation}
where $g^{(0)}_{p}$ is an effective phonon-induced interaction constant in the pairing channel. Here $\mathbf{k}^\prime=\mathbf{k}+\mathbf{q}$ with the phonon momentum $\mathbf{q}$ can be arbitrary in the first Brillouin zone. Although there is no limit in the phonon momentum $\mathbf{q}$, the energy of the phonons is limited approximately by the Debye energy $\hbar \omega_D$, which is much smaller than the Fermi energy $E_F$ in most cases. Therefore, only finite electrons near the Fermi surface can be paired by the phonon-induced attractive pairing interaction $V_1$.  The small energy range of these paired electrons near the Fermi energy leads to low superconducting critical temperature $T_c$. If the pairing energy range can be enlarged, $T_c$ may be enhanced. This is one routine to search for high-$T_c$ superconductors. 

Fundamentally, the microscopic Amp\`ere interaction could make contribution to the formation of the Cooper pairs. Following the idea to derive the phonon-induced effective pairing interaction of Eq. (\ref{eqn5.1}), we can also obtain an effective $\mathbf{A}$-field induced pairing interaction
\begin{equation}
V_{2} = -\frac{1}{N}\sum_{\mathbf{k}\mathbf{k}^\prime} g^{(0)}_{J} c^\dag_{\mathbf{k}\uparrow} c^{\dag}_{-\mathbf{k}\downarrow} c_{-\mathbf{k}^\prime \downarrow} c_{\mathbf{k}^\prime \uparrow} , \label{eqn5.2} 
\end{equation}
where $\mathbf{k}^\prime = \mathbf{k}+\mathbf{q}$ and $g^{(0)}_{J}$ is an effective interaction constant defined in the pairing channel. While $g^{(0)}_{p}$ has positive-value contribution from phonons, $g^{(0)}_J$ has negative-value contribution from photons in the pairing channel of $(\mathbf{k}\uparrow, -\mathbf{k}\downarrow)$. This difference comes from an additional negative factor of $[\widehat{\mathbf{q}}\times(\mathbf{k} +\frac{1}{2}\mathbf{\mathbf{q}})]\cdot [\widehat{\mathbf{q}}\times (-\mathbf{k} - \frac{1}{2}\mathbf{q})]$ in $g_J$ of Eq. (\ref{eqn3.3}). Therefore, the effective microscopic Amp\`ere interaction in the pairing channel makes a destructive contribution to the pairing of the superconducting Cooper pairs. 
 
Consider another pairing channel with $(\mathbf{k}_1\uparrow,\mathbf{k}_2\downarrow)$ where $\mathbf{k}_1=\mathbf{k}$ and $\mathbf{k}_2=-\mathbf{k}\pm \mathbf{2 K}$ with $\mathbf{K}$ near $(0,\pi)$ or $(\pi,0)$. Lee \textit{et al}. have proposed that there is an instability in this pairing density wave channel due to the $U(1)$ spin-liquid gauge fields \citep{PALeePRL2007,PALeePRX2014}. Although theoretically, the electromagnetic $\mathbf{A}$ fields can also lead to a similar attractive interaction for the instability of the pairing density wave states, the pairing interaction is largely reduced by the large photon energy $\hbar \omega(\mathbf{q}) = \hbar c q$. When the $\mathbf{A}$-field momentum is so large as the Fermi momentum $q\simeq k_F\simeq 1/a$ with $a\simeq 10^{-10}\, m$, we have $\hbar \omega(\mathbf{q})\simeq 1000\, eV$. In this case, the ratio of the microscopic Amp\`ere interaction to the phonon-induced electron-electron interaction $\alpha_1$ should be modified into a new one:
\begin{equation}
\alpha_3 =  \frac{E_F\cdot \langle \hbar \omega_\lambda(\mathbf{q}) \rangle}{\langle g_{\lambda}(\mathbf{q})\rangle^2}\cdot \frac{\omega_p^2}{\omega^2(\mathbf{q})} =\alpha_1 \cdot \frac{\omega_p^2}{\omega^2(\mathbf{q})}  . \label{eqn5.3}
\end{equation} 
When $\hbar \omega_p \simeq 10\, eV$ and $\hbar \omega(\mathbf{q})\simeq 1000\, eV$, we have $\alpha_3\simeq 10^{-4} \alpha_1 \ll 1$. For the $\mathbf{A}$ fields to have similar magnitude of contribution to the phonon-induced electron-electron interaction, we should have $\hbar \omega(\mathbf{q}) \lesssim \hbar \omega_p\simeq 10eV$, which would lead to $q \lesssim 10^8\, m^{-1}$. These momenta are much smaller than the Fermi momentum $k_F\simeq 1/a \simeq 10^{10} \, m^{-1}$, which can not allow enough coherent scatterings of the electrons near the Fermi surface to form the coherent pairing density wave states. Therefore, the microscopic Amp\`ere interaction has negligible contribution to the instabilities of the pairing density wave states.  

From the discussion in Sec. \ref{sec3}, we know that, in most cases, the microscopic Amp\`ere interaction has a magnitude at the order of $1\, meV$. Therefore, when considering the processes with the dominant energy scale at about $1\, meV$, we should carefully consider the possibility of the microscopic Amp\`ere interaction as a driving force. In the condensed matter field, the heavy fermion superconductors show many novel physical properties with the energy scales at about or smaller than $1 meV$ \citep{StewartNFLRMP2001,StewartFeSCRMP2011}. Therefore, the roles of the microscopic Amp\`ere interaction in the heavy fermion superconductors should be carefully studied. Consider the warm-blood human beings with a healthy body temperature $37^{\circ} \text{C}$. An increase of the body temperature  $\Delta T = 1^{\circ}\text{C} = 1\, K \simeq 0.1\, meV$ can cause various biochemical reaction processes which may lead to fever, cold and other diseases. Thus, the microscopic Amp\`ere interaction may be the driving force of these biochemical reaction processes of human beings. As the atoms or molecules with heavier elements have large relativistic correction effects \citep{Grant2007,JOHNSON1995255,GrzegorzHeliumPRA2001} and the magnitude of the microscopic Amp\`ere interaction is at the order of the $(v/c)^2$ relativistic corrections, the microscopic Amp\`ere interaction will play important roles in these atoms and molecules.  

At the end of this article, we would remark that the theoretical formalism to use the microscopic Coulomb and Amp\`ere interactions to describe the effects of the electromagnetic interactions in charged systems can only be well-defined in the cases without electromagnetic radiations. When a charged system is in such as a high-temperature charged plasma state with strong electromagnetic radiations, the electromagnetic fields must be introduced explicitly and the whole system should include both the charged particles and the electromagnetic fields to ensure the conservations of the energy and momentum of the whole system. In these cases, the Hamiltonian can be defined as following \citep{Mahan1990}:
\begin{equation}
H = \int d\mathbf{r} \sum_\sigma \widebar{\Psi}_\sigma [\frac{1}{2 m} (\widehat{\mathbf{p}}+e \mathbf{A})^2]  \Psi_\sigma + H_I + V_c + H_{p} . \label{eqn5.4}
\end{equation}
Here $H_I=\sum_{i}(\mathbf{P}_i-q_I\mathbf{A})^2/2M$ is the kinetic energy of the positively charged ions, $V_c$ is the microscopic Coulomb interaction defined by Eq. (\ref{eqn1.1}), and $H_p$ is the photon Hamiltonian defined by 
\begin{equation}
H_p = \sum_{\mathbf{q}\lambda} \hbar \omega_{\mathbf{q}} ( a^\dag_{\mathbf{q}\lambda} a_{\mathbf{q}\lambda} + \frac{1}{2} ) , \label{eqn5.5}
\end{equation}
where $a_{\mathbf{q}\lambda}$ and $a^\dag_{\mathbf{q}\lambda}$ are the annihilation and creation operators of the photons with momentum $\mathbf{q}$ and polarization $\lambda$. 
In Eq. (\ref{eqn5.4}), the transverse $\mathbf{A}$ fields can be quantized as
\begin{equation}
\mathbf{A}(\mathbf{r},t) = \sum_{\mathbf{q}\lambda} \boldsymbol{\gamma}_{\lambda}(\mathbf{q}) e^{i\mathbf{q}\cdot \mathbf{r}} ( a_{\mathbf{q}\lambda} e^{-i\omega_{\mathbf{q}} t} + a^\dag_{-\mathbf{q}\lambda} e^{i\omega_{\mathbf{q}} t} ) , \label{eqn5.6}
\end{equation}
where $\boldsymbol{\gamma}_{\lambda}(\mathbf{q})=\sqrt{\frac{\hbar}{2\epsilon_0 \omega_{\mathbf{q}} \mathcal{V}}} \boldsymbol{\xi}_{\lambda}(\mathbf{q})$ with $\boldsymbol{\xi}_{\lambda}(\mathbf{q})$ being the polarization vector of the $\mathbf{A}$ fields. 

Another remark is that the theoretical formalism we have introduced in this article is based upon a non-relativistic quantum electrodynamics theory. A more exact theoretical formalism can be developed from the relativistic quantum electrodynamics theory, where when the electromagnetic fields are integrated out by using a path integral approach, we can obtain a relativistic effective theory for the interacting electron system. The non-relativistic effective theory can be obtained from this relativistic effective theory, which will give us the relativistic corrections such as the spin-orbit interaction, the relativistic kinetic energy corrections, {\it etc.}, as well as the microscopic Coulomb interaction in the charge density-density channel and the microscopic Amp\`ere interaction in the charge current-current channel. The electronic structures of the atoms can be studied from this non-relativistic effective theory where the fluctuation effects of the electromagnetic fields can be included with well-defined contributions. The path integral approach on the relativistic quantum electrodynamics theory we propose here is different from the theoretical formalisms developed previously \citep{Grant2007}, where the low-order contributions of the electromagnetic fields as well as the low-order relativistic corrections are hardly treated self-consistently in perturbation study.  

In summary, we have revisited the fundamental microscopic electromagnetic interactions in charged systems, the well-known microscopic Coulomb interaction in the charge density-density channel and the dynamical microscopic  Amp\`ere interaction in the charge current-current channel. The latter has an energy scale at the order of $1\, meV$ in most cases. Therefore, the physical, chemical and biological processes at about $1\, meV$ energy scale may involve the microscopic Amp\`ere interaction as a driving force.

\section*{ACKNOWLEDGMENTS}
We thank Prof. Jun Chang and Prof. Yunan Yan for invaluable discussions. This work was supported by the Natural Science Foundation of Shandong Province (Grant No. ZR2023MA015).





%

\end{document}